\documentclass[12pt,a4paper]{article}
\usepackage{latexsym}
\usepackage{epsf}
\makeatletter
\@addtoreset{equation}{section}
\makeatother

\def\unit{\hbox to 3.3pt{\hskip1.3pt \vrule height 7pt width .4pt \hskip.7pt
\vrule height 7.85pt width .4pt \kern-2.4pt
\hrulefill \kern-3pt
\raise 4pt\hbox{\char'40}}}

\def\x{\times}

\def \alfa {2 \pi \alpha^\prime}
\def \ikC {({i}_{\hat k} {\hat C})}
\def \iktildeC {({i}_{\hat k} {\hat {\tilde C}})}
\def \ikchi {({i}_{\hat k} {\hat \chi})}
\def \iktildeN {({i}_{\hat k} {\hat {\tilde N}})}
\def \iktildechi {({i}_{\hat k} {\hat {\tilde \chi}})}
\def \j {{\sf g}} 

\begin{document}

\begin{flushright}
\footnotesize
UG-21/98\\
QMW-PH-98-41 \\
SPIN-98/15 \\
December, $1998$
\normalsize
\end{flushright}

\begin{center}


\vspace{.6cm}
{\LARGE {\bf The Kaluza-Klein Monopole in a Massive IIA Background}}

\vspace{.9cm}

{
{\bf Eduardo Eyras}
\footnote{E-mail address: {\tt E.A.Eyras@phys.rug.nl}}\\
{\it Institute for Theoretical Physics\\
University of Groningen \\
Nijenborgh 4, 9747 AG Groningen, The Netherlands}
}

\vspace{.2cm}

{and}

\vspace{.2cm}

{
{\bf Yolanda Lozano}
\footnote{E-mail address: {\tt Y.Lozano@phys.uu.nl}}\\
{\it Spinoza Institute\\
University of Utrecht\\
Leuvenlaan 4, 3508 TD Utrecht, The Netherlands}
}

\vspace{.2cm}


\vspace{.2cm}

\vspace{.8cm}


{\bf Abstract}

\end{center}

\begin{quotation}

\small

We construct the effective action of the KK-monopole in a
massive Type IIA background. We follow two approaches.
First we construct a massive M-theory KK-monopole from which
the IIA monopole is obtained by double dimensional reduction.
This eleven dimensional monopole contains two isometries: one 
under translations of the Taub-NUT coordinate and the other
under massive transformations of the embedding coordinates.
Secondly, we construct the massive T-duality rules that map the
Type IIB NS-5-brane onto the massive Type IIA KK-monopole. This provides
a check of the action constructed from eleven dimensions.

\end{quotation}

\vspace{1cm}

\newpage

\pagestyle{plain}


\newpage
\section{Introduction}


The eleven-dimensional interpretation of the massive Type IIA
supergravity of Romans \cite{Romans} has devoted a lot of
attention recently.
It was shown in \cite{BDHS} that it is not possible 
to construct a covariant 
supergravity theory in eleven dimensions with cosmological
constant. 
One might assume that the mass arises in the Type IIA theory from
the dimensional reduction
procedure, \`a la Scherk-Schwarz, however a theory was obtained
through this kind of construction in \cite{HLW} with the result
that it was not
Romans' supergravity but a different massive supergravity for which
there is no action.

A massive eleven-dimensional supergravity was proposed
in \cite{BLO} with the peculiarity that it 
can only be formulated
when the eleventh direction is compact, with
Lorentz invariance taking place in the other ten dimensional
coordinates. This feature
circumvents the no-go theorem of \cite{BDHS}. The explicit
eleven-dimensional supergravity action depends on the Killing vector
associated to translations along the eleventh coordinate, and
gives the massive Type IIA supergravity action after a direct 
dimensional reduction along this direction. 

More recently \cite{Hull3}
a proposal for massive M-theory has been given
based on the connection between M-theory and Type IIB and the 
relation between the Scherk-Schwarz reduction of Type IIB and
the reduction of massive Type IIA. It remains an interesting open
problem to make contact with the description of massive 
eleven-dimensional supergravity with a Killing vector.

M-branes propagating in a massive background were constructed in 
\cite{LO,BLO} and referred to generically as massive M-branes.
Their effective actions are described by gauged 
sigma-models in which the Killing isometry is gauged. They contain
as well new couplings proportional to the mass which can be
interpreted in terms of ``massive'' solitons. 

Furthermore, two massive branes in the Type IIA theory
can be obtained from a single massive M-brane \cite{BLO}. 
They are obtained as either direct or double
dimensional reduction of the massive M-brane effective 
action along the space-time coordinate
where the isometry is realized.

In this paper we will construct the effective action of the massive 
Type IIA KK-monopole. 
We will follow two approaches.
First of all, we will obtain the action of the IIA KK-monopole 
by performing a double dimensional reduction in the effective action
of the massive eleven-dimensional KK-monopole. 
As explained in \cite{BEL} the construction of
a massive KK-monopole in eleven dimensions 
is more subtle than that of an ordinary 
brane, since
the monopole is already described by a gauged sigma-model 
in the massless case. 

The massive M-KK-monopole giving
rise to the massive D6-brane after a direct dimensional reduction 
was constructed in \cite{BEL}. There 
it was shown that in order to assure
invariance under massive gauge transformations the gauge field
associated to the Taub-NUT isometry
had to transform proportionally to the
mass. Also, new couplings to the Born-Infeld 
field had to be introduced.

In this eleven-dimensional massive monopole the ``mass
isometry direction''
coincides with the Taub-NUT direction,
and the dimensionally reduced action gives the massive D6-brane.
However, in order to obtain a ten dimensional KK-monopole we are not
interested in eliminating the isometry in the Taub-NUT direction
once a double dimensional reduction is performed.
Therefore we first have to find the worldvolume action of a more
general M-KK-monopole in which the invariance under massive
transformations is achieved by gauging 
an isometry other than the Taub-NUT isometry.
Double dimensional reduction along this 
new isometry direction
will give rise to the action of the massive IIA KK-monopole (see Figure 1).
We present
these actions in sections \ref{massive-MKK}, \ref{MKK-->IIAKK} and
\ref{mIIAKK}.

\vskip 12pt
\begin{figure}[!ht]
\begin{center}
\leavevmode
\epsfxsize=13.5cm
\epsfysize=4cm
\epsffile{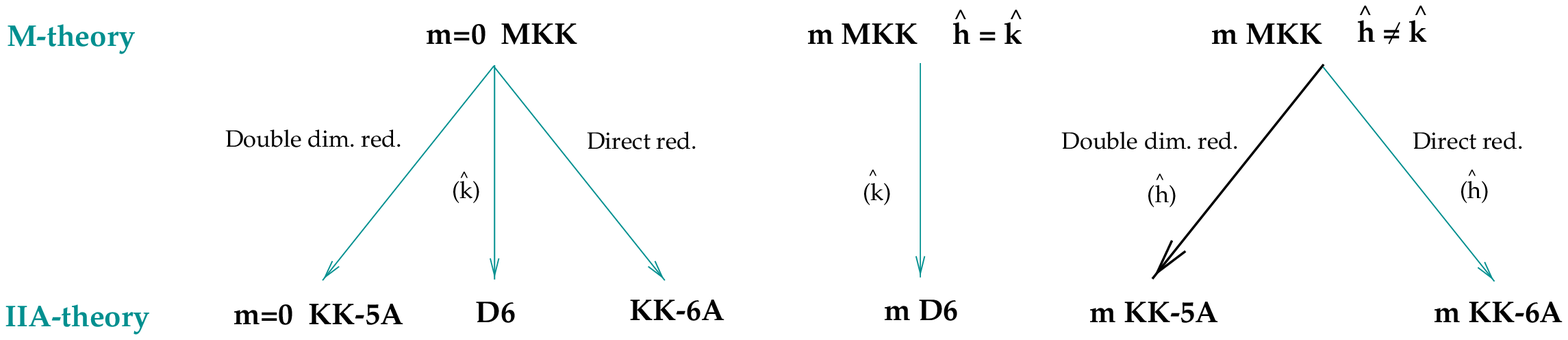}
\caption{\small {\bf Dimensional reductions of the (massive) M-KK-monopole.}
In this figure we display the reductions of the
massless M-KK-monopole
and its two possible massive extensions.
The massless M-KK-monopole is described by a gauged sigma-model
with Killing vector ${\hat k}$. 
Reducing along a worldvolume
coordinate gives rise to the Type IIA KK-monopole, the reduction along
${\hat k}$ gives the D6-brane and the reduction along a transversal
coordinate different than ${\hat k}$ gives a 6-brane (KK-6A)
\cite{U1,U2,U3,MO}, described by a gauged sigma-model. The KK-6A brane
is not associated to a central
charge in the Type IIA supersymmetry algebra
\cite{Hull}.
There are two possible massive 
extensions depending on whether the {\it massive} isometry, ${\hat h}$,
is chosen to be ${\hat h}={\hat k}$ or ${\hat h} \neq {\hat k}$, such that
reducing along ${\hat h}$ gives a massive Type IIA brane.
When ${\hat h}={\hat k}$ we obtain a massive D6-brane.
When ${\hat h}\neq {\hat k}$ we obtain a massive IIA 
KK-monopole (KK-5A in the Figure),
if ${\hat h}$ lies in a worldvolume direction of the M-KK-monopole;
or a massive KK-6A brane, if ${\hat h}$ lies in a transversal
direction.}
\end{center}
\end{figure}

An alternative way to derive the action of the massive IIA KK-monopole
is to start with the action of the IIB NS-5-brane 
constructed in \cite{EJL} and perform a (massive) T-duality
transformation. The massive T-duality rules for the 
target space (dual) potentials that couple to the 5-brane are given in
Appendix A. We present the details of the calculation of the massive IIA
KK-monopole action by this procedure in section 
\ref{IIBNS5-->IIAKK}. The result coincides with that given 
in section \ref{mIIAKK} 
following the double dimensional reduction approach
(see Figure 2).

\vskip 12pt
\begin{figure}[!ht]
\begin{center}
\leavevmode
\epsfxsize= 9cm
\epsfysize= 4cm
\epsffile{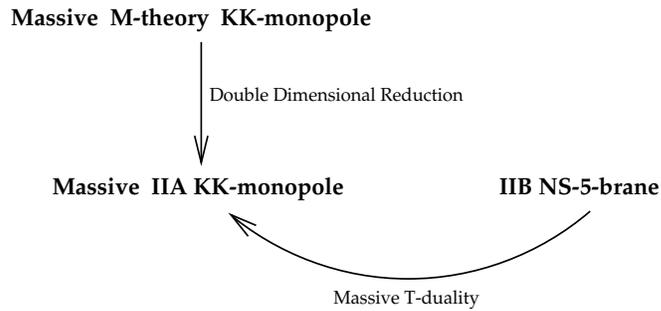}
\caption{\small {\bf Derivation of the massive IIA KK-monopole.}
In this Figure we show the two procedures that we have followed to obtain
the massive IIA KK-monopole. In the first one 
we perform a double dimensional reduction
along the ${\hat h}$ direction of the massive M-KK-monopole
with two gauged isometries, the Taub-NUT ${\hat k}$ and ${\hat h}$.
In the second case, we obtain the same massive
IIA KK-monopole via a massive T-duality transformation on the 
IIB NS-5-brane effective action.}
\label{fig:cuadro1}
\end{center}
\end{figure}


\section{The Massive M-KK-monopole}
\label{massive-MKK}


Let us start by recalling the action of the massless M-KK-monopole constructed
in \cite{BJO} \footnote{For some recent work on KK-monopoles
see \cite{varios}.}.
The KK-monopole in eleven dimensions behaves like a 
6-brane, and its field content is that of the 7-dimensional vector 
multiplet, involving 3 scalars and 1 vector. Since the embedding
coordinates describe $11-7=4$ degrees of freedom one scalar has to be
eliminated by gauging an isometry of the background\footnote{For
constructions of gauged sigma models with WZ term in arbitrary
dimensions see \cite{HS}.}. 

The Taub-NUT
space of the monopole is isometric in its Taub-NUT direction.
Let us denote ${\hat k}$ the Killing vector associated to this 
isometry:

\begin{equation} 
\delta {\hat X}^{\hat \mu}=-{\hat \sigma}^{(0)}{\hat k}^{\hat \mu}\, ,
\end{equation}

\noindent such that the Lie derivatives of all target space 
fields and gauge
parameters with respect to ${\hat k}$ vanish. 
The effective action of the monopole is constructed by replacing 
ordinary derivatives by covariant derivatives:

\begin{equation}
\partial_{\hat \imath}{\hat X}^{\hat \mu}\rightarrow D_{\hat \imath}
{\hat X}^{\hat \mu}=\partial_{\hat \imath}{\hat X}^{\hat \mu}+
{\hat A}_{\hat \imath}{\hat k}^{\hat \mu}\, ,
\end{equation}

\noindent with ${\hat A}_{\hat \imath}$ a dependent gauge field
given by: 

\begin{equation}
{\hat A}_{\hat \imath}=|{\hat k}|^{-2}\partial_{\hat \imath}
{\hat X}^{\hat \mu}{\hat k}_{\hat \mu}\, ,
\end{equation}

\noindent where $|{\hat k}|^2=-{\hat k}^{\hat \mu}
{\hat k}^{\hat \nu}{\hat g}_{{\hat \mu}{\hat \nu}}$.
Accordingly, the gauge transformation of ${\hat A}_{\hat \imath}$ is:

\begin{equation}
\delta {\hat A}_{\hat \imath}=\partial_{\hat \imath}
{\hat \sigma}^{(0)}\, .
\end{equation}

\noindent In this way the effective metric becomes:

\begin{equation}
{\hat g}_{{\hat \mu}{\hat \nu}}D_{\hat \imath}
{\hat X}^{\hat \mu}D_{\hat \jmath}{\hat X}^{\hat \nu}
={\hat \Pi}_{{\hat \mu}{\hat \nu}}\partial_{\hat \imath}
{\hat X}^{\hat \mu}\partial_{\hat \jmath}
{\hat X}^{\hat \nu}\, ,
\end{equation}

\noindent with 

\begin{equation}
{\hat \Pi}_{{\hat \mu}{\hat \nu}}=
{\hat g}_{{\hat \mu}{\hat \nu}}+|{\hat k}|^{-2}
{\hat k}_{\hat \mu}{\hat k}_{\hat \nu}\, .
\end{equation}

\noindent Since ${\hat \Pi}_{{\hat \mu}{\hat \nu}}
{\hat k}^{\hat \nu}=0$ the isometry direction is effectively
eliminated from the action.

The construction of the ${\hat \sigma}^{(0)}$--invariant WZ term has
been given in \cite{BEL}.
We summarize the target space and worldvolume field content in
Tables \ref{table1} and \ref{table2}.

\begin{table}[!ht]
\renewcommand{\arraystretch}{1.5}
\begin{center}
\begin{tabular}{|c|c|}
\hline
Target space & Gauge     \\
Field        & Parameter \\
\hline\hline
${\hat g}_{{\hat \mu}{\hat \nu}}$
& \\
\hline
${\hat C}_{{\hat \mu}{\hat \nu}{\hat \rho}}$ &
 ${\hat \chi}_{{\hat \mu}{\hat \nu}}$ \\
\hline
$(i_{\hat k} {\hat C})_{{\hat \mu}{\hat \nu}}$&
$(i_{\hat k} {\hat \chi})_{{\hat \mu}}$\\
\hline
${\hat {\tilde C}}_{{\hat \mu}_1 \dots {\hat \mu}_6}$
& ${\hat {\tilde \chi}}_{{\hat \mu}_1 \dots {\hat \mu}_5}$ \\
\hline
$\iktildeC_{{\hat \mu}_1 \dots {\hat \mu}_5}$
& $\iktildechi_{{\hat \mu}_1 \dots {\hat \mu}_4}$ \\
\hline
${\hat N}_{{\hat \mu}_1 \dots {\hat \mu}_8}$ 
& ${\hat \Omega}_{{\hat \mu}_1 \dots {\hat \mu}_7}$ \\
\hline
$\iktildeN_{{\hat \mu}_1 \dots {\hat \mu}_7}$ & 
 $(i_{\hat k} {\hat \Omega})_{{\hat \mu}_1 \dots {\hat \mu}_6}$\\
\hline
\end{tabular}
\end{center}  
\caption{\label{table1} \small {\bf Target space fields in the 
M-KK-monopole.}
This table shows the 11-dimensional target space
fields that couple to the M-KK-monopole, 
together with their gauge parameters. We also include
the contractions with the Killing vector ${\hat k}^{\hat \mu}$.
The field ${\hat {\tilde C}}$ is the Poincar\'e dual of ${\hat C}$,
the 3-form of eleven-dimensional supergravity,
and ${\hat N}$ is the Poincar\'e dual of the Killing vector, 
considered
as a 1-form ${\hat k}_{\hat \mu}$.
}
\renewcommand{\arraystretch}{1}
\end{table}

\begin{table}[h]
\renewcommand{\arraystretch}{1.5}
\begin{center}
\begin{tabular}{|c|c|c|c|}
\hline
Worldvolume     & $\sharp$ of   \\      
Field           & d.o.f         \\
\hline\hline
${\hat X}^{{\hat \mu}}$ & $11 - 7 - (1) = 3$ \\
\hline
$ {\hat \omega}^{(1)}_{{\hat \imath}}$ &
 $7 - 2 = 5$\\
\hline
${\hat \omega}^{(6)}_{{\hat \imath}_1 \dots {\hat \imath}_6}$
& $-$ \\
\hline
\end{tabular}
\end{center}  
\caption{\small \label{table2} {\bf Worldvolume fields.}
In this table we summarize the
worldvolume fields, together with their
degrees of freedom, that occur in the worldvolume
action of the M--theory KK--monopole. 
The worldvolume scalars ${\hat X}^{\hat \mu}$ are the embedding coordinates,
${\hat \omega}^{(1)}$ is a 1-form and
${\hat \omega}^{(6)}$ is a non propagating 6-form that describes the 
tension of the monopole. Due to the gauging the embedding scalars
describe 3 and not 4 degrees of freedom as indicated in the table.
}
\renewcommand{\arraystretch}{1}
\end{table}

The resulting effective action gives the D6-brane of the Type
IIA theory after a direct dimensional reduction along the Taub-NUT
direction \cite{BJO}.

In \cite{BEL} the effective action of the M-KK-monopole giving
rise to the D6-brane of the massive Type IIA theory was also
constructed. As discussed in \cite{BLO} eleven-dimensional
massive branes are described by gauged sigma models with gauge
coupling constant proportional to $m$. The massive D6-brane is
obtained by reducing the massive KK-monopole in which the two
Killing vectors associated to the mass and the Taub-NUT 
isometries coincide. The gauge field ${\hat A}$ must be attributed
mass transformation rules and extra terms need to be added to the
WZ part. As we mentioned in the Introduction this massive 
M-KK-monopole cannot give rise to the massive Type IIA KK-monopole
after a double dimensional reduction, since the gauged isometry
disappears in the reduction. However it is possible to construct
a massive M-KK-monopole in which the two isometry directions 
associated to the mass and the Taub-NUT space are different. 
Double dimensionally reducing along the direction associated to the
mass will give rise to the effective action of the massive IIA
KK-monopole.

It was shown in \cite{BLO} that a massive brane in eleven dimensions is
obtained by gauging an isometry generated by a Killing vector ${\hat h}$:

\begin{equation}
\delta_{{\hat \rho}^{(0)}}{\hat X}^{\hat \mu}=\frac{m}{2}
l_p^2 \, {\hat \rho}^{(0)}{\hat h}^{\hat \mu}\, ,
\end{equation}

\noindent through the introduction of a gauge field 
${\hat b}_{\hat \imath}$ transforming as\footnote{Here
$l_p$ is the eleven-dimensional Planck length.}:

\begin{equation}
\delta {\hat b}_{\hat \imath}=\partial_{\hat \imath}
{\hat \rho}^{(0)}- l_p^{-2} \, 
(i_{\hat h}{\hat \chi})_{\hat \imath}\, ,
\end{equation}

\noindent where $(i_{\hat h}{\hat \chi})$ is (the pull-back of)
the interior product of the gauge parameter of the 
eleven-dimensional 3-form ${\hat C}$ with the Killing vector
${\hat h}$. Substituting ordinary derivatives by covariant
derivatives

\begin{equation}
{\cal D}_{\hat \imath}{\hat X}^{\hat \mu}=\partial_{\hat \imath}
{\hat X}^{\hat \mu}-\frac{m}{2} l_p^2 \,
{\hat b}_{\hat \imath}{\hat h}^{\hat \mu}
\end{equation}

\noindent one assures invariance under massive transformations:

\begin{equation}
\label{massivetr}
\delta_{\hat \chi}{\hat L}_{{\hat \mu}_1\dots {\hat \mu}_r}=
r\frac{m}{2} (-1)^r (i_{\hat h}{\hat \chi})_{[{\hat \mu}_1}
(i_{\hat h}{\hat L})_{{\hat \mu}_2\dots {\hat \mu}_r]}\, ,
\end{equation}

\noindent for a rank $r$ 11 dim form ${\hat L}$, and

\begin{equation}
\delta_{\hat \chi}{\hat g}_{{\hat \mu}{\hat \nu}}=-m
(i_{\hat h}{\hat \chi})_{({\hat \mu}}
(i_{\hat h} {\hat g})_{{\hat \nu})}\, ,
\end{equation}

\noindent for the 11 dim metric. 

These transformations give rise to the known
massive transformations of the Type IIA background fields 
after dimensional reduction.
Together with the gauging it is 
necessary to include additional ${\hat b}$--dependent terms
(proportional to the mass) to achieve invariance under massive
transformations. 

This suggests that we should construct the massive M-KK-monopole
by substituting:

\begin{equation}
{\hat \Pi}_{{\hat \mu}{\hat \nu}}\partial_{\hat \imath}
{\hat X}^{\hat \mu}\partial_{\hat \jmath}{\hat X}^{\hat \nu}
\rightarrow {\hat \Pi}_{{\hat \mu}{\hat \nu}}
{\cal D}_{\hat \imath}{\hat X}^{\hat \mu}{\cal D}_{\hat \jmath}
{\hat X}^{\hat \nu}\, ,
\end{equation}

\noindent with ${\cal D}{\hat X}$ as defined above. This can 
also be written as:

\begin{equation}
{\hat \Pi}_{{\hat \mu}{\hat \nu}}{\cal D}_{\hat \imath}
{\hat X}^{\hat \mu}{\cal D}_{\hat \jmath}{\hat X}^{\hat \nu}
={\hat g}_{{\hat \mu}{\hat \nu}}D_{\hat \imath}
{\hat X}^{\hat \mu}D_{\hat \jmath}{\hat X}^{\hat \nu}\, ,
\end{equation}

\noindent with

\begin{equation}
\label{covdev}
D_{\hat \imath}{\hat X}^{\hat \mu}\equiv 
\partial_{\hat \imath}{\hat X}^{\hat \mu}
+{\hat A}_{\hat \imath}{\hat k}^{\hat \mu}
-\frac{m}{2}l_p^2 \, {\hat b}_{\hat \imath}
{\hat h}^{\hat \mu}\, ,
\end{equation}

\noindent and:

\begin{equation}
\label{Atrans}
{\hat A}_{\hat \imath}=|{\hat k}|^{-2}{\hat k}_{\hat \mu}
\left( \partial_{\hat \imath}{\hat X}^{\hat \mu}-\frac{m}{2} \right.
\left. l_p^2 \, {\hat b}_{\hat \imath}{\hat h}^{\hat \mu}\right)\, .
\end{equation}

\noindent ${\hat \Pi}_{{\hat \mu}{\hat \nu}}$ transforms as a
metric under massive transformations iff

\begin{equation}
\label{vectortr}
\delta {\hat k}_{\hat \mu}=-\frac{m}{2}
(i_{\hat h}{\hat \chi})_{\hat \mu}(i_{\hat h}{\hat k})\, ,
\end{equation}

\noindent which implies that the Killing vector associated to the
Taub-NUT isometry must be attributed a massive transformation:

\begin{equation}
\label{ktrans}
\delta {\hat k}^{\hat \mu}=\frac{m}{2}(i_{\hat k}i_{\hat h}
{\hat \chi}){\hat h}^{\hat \mu}\, .
\end{equation}

\noindent The transformation rule (\ref{vectortr}) is that of a 
vector under massive transformations (see (\ref{massivetr})).
We showed in \cite{BEL} that ${\hat k}_{\hat \mu}$ had to be
considered as a target space 1-form in order to construct the
action of the 11 dim KK-monopole. In particular, the monopole is
charged with respect to its dual 8-form\footnote{To be precise,
with respect to its interior product with the Killing vector
${\hat k}$.}. We see here that this is also the case regarding
massive transformations. With this transformation rule the interior
product of ${\hat k}$ with any eleven-dimensional $r$-form 
transforms according to (\ref{massivetr}) (see Appendix B).
 
The fact that ${\hat k}^{\hat \mu}$ transforms under massive
transformations implies that the Killing condition and the
massive transformations do not commute\footnote{They commute
if $(i_{\hat k}i_{\hat h}{\hat \chi})=0$, which is not the
most general case. Note, however, that if in ten dimensions we
set the component of the RR 1-form along the Taub-NUT 
direction to zero, which can always be done since $C^{(1)}$
is a non-physical Stueckelberg field \cite{Romans,BRGPT}, then
this condition is satisfied.}.

In the system of adapted coordinates to the isometry generated
by ${\hat h}$: ${\hat h}^{\hat \mu}=\delta^{\hat \mu}{}_y$, we
can define:

\begin{equation}
{\hat k}^y=\frac{m}{2} l_p^2 \, {\hat \omega}^{(0)}\, ,
\end{equation}

\noindent with

\begin{equation}
\delta {\hat \omega}^{(0)}=\frac{1}{2\pi\alpha^\prime}
(i_{\hat k}i_{\hat h}{\hat \chi})\, .
\end{equation}

\noindent In the reduction to ten dimensions ${\hat k}^{\mu}$ 
(with ${\mu}$ a ten dimensional index) will be the Killing 
vector generating translations along the Taub-NUT direction, 
and ${\hat \omega}^{(0)}$ an extra worldvolume scalar 
needed to compensate for certain massive transformations.

It is easy to check that the invariance under ${\hat \sigma}^{(0)}$
is preserved\footnote{In the kinetic term. We will comment
later on the WZ term.} if the ${\hat b}$ field transforms as:

\begin{equation}
\label{btrans}
\delta {\hat b}_{\hat \imath}=
\partial_{\hat \imath} {\hat \rho}^{(0)}
-{\hat \sigma}^{(0)}\partial_{\hat \imath}
{\hat \omega}^{(0)}
- l_p^{-2} \, (i_{\hat h}{\hat \chi})_{\hat \imath}\, .
\end{equation}

The gauge transformation rule of ${\hat A}_{\hat \imath}$ (given 
by (\ref{Atrans})) is still $\delta {\hat A}_{\hat \imath}=
\partial_{\hat \imath}{\hat \sigma}^{(0)}$, as in the massless case.
                              
The action that we propose for the massive M-theory KK-monopole 
is the following:

\begin{eqnarray}
\label{MKKaction}
{\hat S}&=&-T_{{\rm mMKK}}\int d^7 {\hat \xi} {\hat k}^2\sqrt{|{\rm det}
(D_{\hat \imath}{\hat X}^{\hat \mu}D_{\hat \jmath}
{\hat X}^{\hat \nu}{\hat g}_{{\hat \mu}{\hat \nu}}+
 l_p^2 |{\hat k}|^{-1}{\hat {\cal K}}^{(2)}_{{\hat \imath}
{\hat \jmath}})|} 
\nonumber \\
&&+\,\, \frac{1}{7!} l_p^2 \, T_{{\rm mMKK}}\int d^7 {\hat \xi}
\,\, \varepsilon^{{\hat \imath}_1\dots{\hat \imath}_7}
{\hat {\cal K}}^{(7)}_{{\hat \imath}_1\dots {\hat \imath}_7}\, .
\end{eqnarray}

\noindent The covariant derivatives are defined by (\ref{covdev})
and ${\hat {\cal K}}^{(2)}$ 
is the massive field strength of the worldvolume field
${\hat \omega}^{(1)}$:

\begin{equation}
{\hat {\cal K}}^{(2)}=2\partial {\hat \omega}^{(1)}+
l_p^{-2} \, D{\hat X}^{\hat \mu}D
{\hat X}^{\hat \nu}(i_{\hat k}{\hat C})_{{\hat \mu}{\hat \nu}}
-m l_p^2 \, \partial{\hat \omega}^{(0)}{\hat b}  \, .
\end{equation}

\noindent Finally, ${\hat {\cal K}}^{(7)}$ is given by:

\begin{equation}
\begin{array}{rcl}
{\hat {\cal K}}^{(7)} &=&
7 \left\{ \partial {\hat \omega}^{(6)}
+m {\hat \omega}^{(7)}
-\frac32  l_p^2 \, m{\hat d}^{(5)}
\left(2\partial{\hat\omega}^{(1)}
-ml_p^2 \,\partial {\hat \omega}^{(0)}{\hat b} \right) \right. \\
& & \\
& &
\left.  - {1 \over 7} l_p^{-2} \, 
D {\hat X}^{{\hat \mu}_1} \dots D
 {\hat X}^{{\hat \mu}_7}
(i_{\hat k} {\hat N})_{{\hat \mu}_1 \dots {\hat \mu}_7}
+3 D  {\hat X}^{{\hat \mu}_1} \dots
D  {\hat X}^{{\hat \mu}_5}
 (i_{\hat k}{\hat {\tilde C}})_{{\hat \mu}_1 \dots {\hat \mu}_5}
{\hat {\cal K}}^{(2)}\right.
\\ & & \\
& &
\left. - {5} l_p^{-2} \, D  {\hat X}^{{\hat \mu}_1}
\dots D  {\hat X}^{{\hat \mu}_7}
{\hat C}_{{\hat \mu}_1 \dots {\hat \mu}_3} 
\ikC_{{\hat \mu}_4 {\hat \mu}_5}
 \ikC_{{\hat \mu}_6 {\hat \mu}_7} \right.\\
& &\\
& &
\left. -15 D {\hat X}^{{\hat \mu}_1} \dots
D {\hat X}^{{\hat \mu}_5}
{\hat C}_{{\hat \mu}_1 \dots {\hat \mu}_3} \ikC_{{\hat \mu}_4 {\hat \mu}_5}
 \left( 2\partial {\hat \omega}^{(1)} 
-m l_p^2 \, \partial{\hat \omega}^{(0)}{\hat b}\right)\right.
\\ & & \\
& &
\left. -60 l_p^2 \,  D {\hat X}^{{\hat \mu}_1} \dots
D {\hat X}^{{\hat \mu}_3} {\hat C}_{{\hat \mu}_1 \dots
{\hat \mu}_3}\partial {\hat \omega}^{(1)}
\left( \partial {\hat \omega}^{(1)}- 
m l_p^2 \, \partial {\hat \omega}^{(0)}{\hat b} \right)\right.\\
& &\\
& &
\left. -60 l_p^4 \,  {\hat A} \left( 2\partial {\hat \omega}^{(1)}
-3m l_p^2 \, \partial{\hat \omega}^{(0)}{\hat b} \right)
\partial {\hat \omega}^{(1)}\partial {\hat \omega}^{(1)}
 \right\} \, . \\
\end{array}
\end{equation}

\noindent This action is invariant under the gauge transformations given in
Appendices B and \ref{massiveMKK}. 
Invariance under ${\hat \rho}^{(0)}$ transformations is
assured by the presence of covariant derivatives. It can also be seen that
with the transformation rule (\ref{btrans}) for the ${\hat b}$ field
the action is 
invariant under ${\hat \sigma}^{(0)}$ transformations. 

Notice that we have introduced a new auxiliary field, 
${\hat d}^{(5)}$, associated to the dual massive transformations
with parameter $(i_{\hat h}{\hat \Sigma})$ (see \cite{BLO}).
This field transforms proportionally to 
$(i_{\hat k}i_{\hat h} {\hat \Sigma})$ (see Appendix C.1), since
only contractions with the Taub-NUT Killing vector need to be
cancelled in the KK-monopole action.

Moreover, ${\hat b}$ is a non-propagating field in this action, playing
the role of gauge field for the ${\hat h}$ isometry\footnote{Additional
$m{\hat b}$ couplings are necessary as well in the WZ term to cancel
certain gauge variations.}.
We will see in the next section that it disappears in the double
dimensionally reduced action. This means that there are no fundamental
strings ending on the monopole.

When ${\hat k}$ and ${\hat h}$ are parallel, as in \cite{BEL}, 
${\hat \omega}^{(1)}$ transforms  
like ${\hat b}$ (in (\ref{btrans})) \footnote{In this case 
${\hat \omega}^{(0)}=0$.}
and it is not necessary to 
introduce this new field.
Furthermore, an
extra term $-15m l_p^6 \, {\hat b}\partial {\hat b}\partial {\hat b}
\partial {\hat b}$ in the
WZ part of the action accounts for the variation of the ${\hat
  \omega}^{(7)}$ field\footnote{With the
  modification of the transformation law of ${\hat \omega}^{(6)}$ 
  found in \cite{BEL}.}, which is also not needed.
This is also the case for ${\hat \omega}^{(0)}$ and 
${\hat d}^{(5)}$.

In Table \ref{table-mMKK} we summarize the worldvolume fields
present in (\ref{MKKaction}). All these fields can be given an
interpretation in terms of solitons on the KK-monopole. 

\begin{table}[h]
\renewcommand{\arraystretch}{1.5}
\begin{center}
\begin{tabular}{|c|c|}
\hline
Worldvolume     & Field  \\
Field           & Strength  \\
\hline\hline
${\hat \omega}^{(1)}_{{\hat \imath}}$ &
 ${\hat {\cal K}}^{(2)}_{{\hat \imath}{\hat \jmath}}$ \\
\hline
${\hat d}^{(5)}_{{\hat \imath}_1{\hat \imath}_2}$ & 
$ $  \\
\hline
${\hat \omega}^{(6)}_{{\hat \imath}_1 \dots {\hat \imath}_6}$
& ${\hat {\cal K}}^{(7)}_{{\hat \imath}_1 \dots {\hat \imath}_7}$ \\
\hline
${\hat \omega}^{(7)}_{{\hat \imath}_1 \dots {\hat \imath}_7}$ 
 & \\
\hline
\end{tabular}
\end{center}
\caption{\label{table-mMKK} \small {\bf Worldvolume field content of the
massive M-theory KK-monopole.} In this table we give the
worldvolume fields, together with their field strengths,
present in the worldvolume action of the massive M--theory KK--monopole.}
\renewcommand{\arraystretch}{1}
\end{table}

We find, as in the massless case, a 1-form ${\hat \omega}^{(1)}$,
describing a 0-brane soliton in the worldvolume of the KK-monopole.
Its dual 4-form describes a 3-brane soliton. 
They correspond to the intersections: $(0|{\rm M}2,{\rm MKK})$ 
and $(3|{\rm M}5,{\rm MKK})$,
respectively. 
The monopole contains as well a 
4-brane soliton which couples to the 5-form dual to one of the embedding
scalars: $(4|{\rm MKK},{\rm MKK})_{1,2}$. 
All these intersections have been discussed in \cite{BREJS,BGT}.

In the massive case the field ${\hat d}^{(5)}$ couples to the 5-brane
soliton represented by the configuration\footnote{We use here a 
notation where $\times (-)$ indicates
  a worldvolume (transverse) direction. The first $\times$ in a
  row indicates the time direction.} \cite{deRoo,BGT}:

\begin{displaymath}
 (5|{\rm M}9,{\rm MKK})=
\mbox{ 
$\left\{ \begin{array}{c|cccccccccc}
         \x & \x & \x  & \x & \x & z & - & \x & \x  & \x  & \x \\
         \x & \x & \x   & \x  & \x & \x & \x & z & -  & - & -           
                                    \end{array} \right.$}   
\end{displaymath}

\noindent Here the $z$-direction in the monopole corresponds to the
isometry direction of the Taub-NUT space. A single M9-brane
contains as well a Killing isometry in its worldvolume, as has been
discussed in \cite{BvdS,BEHHLS,proci}. This 
isometric direction has been depicted as well as a $z$-direction.
The 5-brane soliton predicted by the M-KK and M9-brane worldvolume
supersymmetry algebras \cite{BGT} is realized as a 4-brane
soliton given that it cannot develop a worldvolume direction along the
isometry of the M9-brane. This is in agreement with the worldvolume
field content that we have found for the massive M-KK-monopole, since
the only worldvolume field to which this soliton can couple
is the 5-form ${\hat d}^{(5)}$.

The 6-form ${\hat \omega}^{(6)}$ is interpreted as the tension of
the monopole and couples to the 5-brane soliton
realized as the embedding of an M5-brane on the
KK-monopole \cite{Tsey,BREJS}.
In the massive case it also
plays the role of Stueckelberg field for the auxiliary field
${\hat \omega}^{(7)}$ (see their ${\hat \rho}^{(6)}$ 
transformation rules in Appendix C.1).

Finally, ${\hat \omega}^{(7)}$ couples to the 6-brane soliton describing
the embedding of the monopole in an M9-brane: $(6|{\rm M}9,{\rm MKK})$.
The M9-brane contains a 1-form vector field in its worldvolume
\cite{proci}. The dual of this massive 1-form in the nine
dimensional worldvolume is a 7-form field which from the point of view
of the M9-brane is the worldvolume field that couples to the
6-brane soliton.

  
\section{Double Dimensional Reduction:\\
Massive MKK $\rightarrow$ Massive KK-5A}
\label{MKK-->IIAKK}


We can now proceed and perform the double dimensional reduction
of the action constructed in the previous section.
We will see that the Killing isometry associated to translations of the
Taub-NUT coordinate is restored in this process.

In adapted coordinates 
${\hat h}^{\hat \mu}=\delta^{\hat \mu}{}_y$, we take the ansatz
for the worldvolume reduction:
\begin{equation}
{\hat X}^y=Y={\hat \xi}^6
\end{equation}

\noindent with all other worldvolume fields and gauge parameters
independent of ${\hat \xi}^6$. 

The ${\hat k}$ transformation rule (\ref{ktrans}) implies that 
this vector gets a $y$-component under massive gauge transformations.
Therefore we reduce it as:

\begin{equation}
{\hat k}^\mu=k^\mu \, ,\qquad
{\hat k}^y = m 2\pi\alpha^\prime \omega^{(0)} \, .
\end{equation}

\noindent $k^\mu$ will be the Killing vector associated to the isometry 
of the Taub-NUT space of the IIA KK-monopole.

In order to keep track of all the
gauge transformations we have to introduce a compensating gauge
transformation:

\begin{equation}
\delta {\hat \xi}^{{\hat \imath}}=\delta^{{\hat \imath}6}
[-\Lambda^{(0)}+\frac{m}{2}(2\pi\alpha^\prime)
{\rho}^{(0)}-\frac{m}{2}(2\pi\alpha^\prime)\sigma^{(0)}
\omega^{(0)}]\, ,
\end{equation}
\noindent where $\Lambda^{(0)}$ is a g.c.t. in the direction $Y$.

The reduction rules for the background fields and gauge parameters
can be found for instance in \cite{BEL}.
The worldvolume fields reduce as:

\begin{equation}
\begin{array}{rclrcl}
l_p^2 \, {\hat b}_i&=& \alfa \, b_i \, ,&
 l_p^2 \, {\hat b}_6 &=& \alfa \, v^{(0)} \, ,\\
& & \\
l_p^2 \, {\hat \omega}^{(1)}_i &=& \alfa \, {\omega}^{(1)}_i\, ,& \,\,\,\,\,\,
l_p^2 \, {\hat \omega}^{(1)}_6 &=& \alfa \, \omega^{(0)} \, ,\\
& & \\
l_p^2 \, {\hat d}^{(5)}_{i_1\dots i_5}&=& \alfa \, d^{(5)}_{i_1\dots i_5} 
\, , & \,\,\,\,\,\,
l_p^2 \, {\hat \omega}^{(6)}_{i_1\dots i_5 6}&=&
\alfa \, \omega^{(5)}_{i_1\dots i_5} \, ,\\
\end{array}
\end{equation}

\begin{displaymath}
l_p^2 \, {\hat \omega}^{(7)}_{i_1\dots i_6 6}=
-\frac37 (2\pi\alpha^\prime)^2 \, v^{(0)}
\partial_{[i_1}\omega^{(5)}_{i_2\dots i_6]} 
+\frac67 (\alfa) \left(1-\frac{m}{2}(2\pi\alpha^\prime) v^{(0)}\right)
 \, \omega^{(6)}_{i_1\dots i_6} \, .
\end{displaymath}

The (modified) gauge transformations of the new, reduced, worldvolume 
fields can be found in Appendix C.2.

The reduction of ${\hat \omega}^{(6)}$ holds up to a total 
derivative (see Appendix C). 
The reduction of ${\hat d}^{(5)}_{i_1\dots i_4 6}$ gives a
worldvolume 4-form which is gauge invariant and contributes to the reduced
action with a decoupled term, therefore we have fixed it to zero. The scalar
field $v^{(0)}$ also has vanishing transformation law.
However this field contributes to the double dimensionally reduced
action as:

\begin{equation}
\int d^6\xi (1-\frac{m}{2}(2\pi\alpha^\prime)v^{(0)})
{\cal L}_{{\rm mAKK}}\, ,
\end{equation}

\noindent and is constrained to a constant by
the equation of motion of the worldvolume field $\omega^{(5)}$
playing the role of tension of the IIA monopole. In general it modifies
the tension of the new massive p-brane as \cite{BLO}: 

\begin{equation}
\label{modifi}
T_{\rm mIIA} = \left( 1 - {m \over 2} (\alfa)v^{(0)} \right) 
T_{\rm mM} \, .
\end{equation} 

This has the implication that for the particular value
$v^{(0)}=2/m(2\pi\alpha^\prime)$ the brane tension vanishes.
The physical mechanism behind this phenomenon is unclear and
it would be interesting to investigate (see the Conclusions for a
further discussion).

The dependent gauge field ${\hat A}$ reduces as:

\begin{eqnarray}
{\hat A}_i&=&\left( 1+e^{2\phi}|k|^{-2}(i_k C^{(1)}+
m\pi\alpha^\prime\omega^{(0)})^2 \right)^{-1}\times\nonumber\\
&&\times\left(A_i-e^{2\phi}
|k|^{-2}(C^{(1)}_i-m\pi\alpha^\prime b_i)(i_kC^{(1)}+
m\pi\alpha^\prime\omega^{(0)}) \right)\, ,
\end{eqnarray}

\noindent where $A_i\equiv |k|^{-2}k_{\mu}\partial_i X^\mu$, and

\begin{eqnarray}
{\hat A}_6&=&- \left( 1+e^{2\phi}|k|^{-2}(i_k C^{(1)}+m\pi\alpha^\prime
\omega^{(0)})^2 \right)^{-1}\times\nonumber\\
&&\times \,\, e^{2\phi}
|k|^{-2}(1-m\pi\alpha^\prime v^{(0)})(i_k C^{(1)}+
m\pi\alpha^\prime\omega^{(0)}) \, .
\end{eqnarray}


\section{The Action of the Massive IIA KK-monopole}
\label{mIIAKK}


The double dimensional reduction of the action of the massive M-KK-monopole 
gives: 

\begin{eqnarray}
\label{accionmasiva}
S&=& 
-T_{\rm mAKK} \int d^6\xi \,\, k^2
e^{-2\phi}
\sqrt{1+e^{2\phi}k^{-2}(i_k C^{(1)}+m\pi\alpha^\prime\omega^{(0)})^2}
\times
\nonumber \\
&&\hspace{-1.5cm}
\times \sqrt{\Biggl|{\rm det} \left( \Pi_{ij}
-(2\pi\alpha^\prime)^2
k^{-2}{\cal K}^{(1)}_i {\cal K}^{(1)}_j
+\frac{(2\pi\alpha^\prime)k^{-1}e^{\phi}}{\sqrt{1+e^{2\phi}
k^{-2}(i_k C^{(1)}+m\pi\alpha^\prime\omega^{(0)})^2}}
{\cal K}^{(2)}_{ij} \right) \Biggr|}\nonumber \\
&&+\,\, \frac{1}{6!}(2\pi\alpha^\prime)T_{\rm mAKK} \int d^6\xi 
\epsilon^{i_1\dots i_6} 
{\cal K}^{(6)}_{i_1\dots i_6}\, .
\end{eqnarray}

\noindent The covariant derivative is defined as:
$D_i X^\mu=\partial_i X^\mu+A_i k^\mu$ and
\begin{equation}
\Pi_{ij}  =D_i X^{\mu}D_j X^{\nu}g_{\mu\nu} \, .
\end{equation}

\noindent The tension $T_{\rm mAKK}$ is the modified tension
given by (\ref{modifi}).

The gauge invariant forms ${\cal K}^{(2)}$ and ${\cal K}^{(1)}$ 
are the field strengths of
$\omega^{(1)}$ and $\omega^{(0)}$, respectively:

\begin{equation}
\label{curvas}
\begin{array}{rcl}
{\cal K}^{(2)} &=& 2\partial \omega^{(1)}+
\frac{1}{2\pi\alpha^\prime}(i_k C^{(3)})
- 2 {\cal K}^{(1)} (DX C^{(1)})
\\& &\\& &
+{m \over 2} \omega^{(0)} (DXDX B) - m (\alfa) \omega^{(0)} 
\partial \omega^{(0)} A \, ,\\ & &\\
{\cal K}^{(1)} &=& \partial {\omega}^{(0)}-\frac{1}{2\pi\alpha^\prime}
(i_k B) \, ,\\
\end{array}
\end{equation}

\noindent and the WZ term is the field strength of the worldvolume field
$\omega^{(5)}$, playing the role of tension of the IIA KK-monopole:
 
\begin{equation}
\begin{array}{rcl}
\label{WZKK}
{\cal K}^{(6)} &=& 6 (\partial \omega^{(5)}+m\omega^{(6)})
-3m(2\pi\alpha^\prime)d^{(5)}\partial\omega^{(0)}+
\frac{1}{2\pi\alpha^\prime}(i_k N)
\\
& & \\ & &
-15 (i_k C^{(5)})(2\partial\omega^{(1)}
+\frac{1}{2\pi\alpha^\prime}(i_k C^{(3)})+\frac{m}{2}\omega^{(0)}B)
\\  
& & \\ & &
-6((i_k {\tilde B})
-\frac{m}{2}(2\pi\alpha^\prime)\omega^{(0)}C^{(5)})
{\cal K}^{(1)} \\ 
& & \\ & &
 -60 (\alfa) DX^{{\mu}} DX^{{\nu}} DX^{{\rho}} 
C^{(3)}_{{\mu} {\nu} {\rho}} 
\, {\cal K}^{(1)} {\cal K}^{(2)}
\\ & & \\ & & 
+{30 \over 2\pi\alpha^{\prime}}B
 (i_k C^{(3)})^2
+30DX^{\mu} DX^{\nu} DX^{\rho} C^{(3)}_{\mu \nu \rho} 
( i_k C^{(3)}) {\cal K}^{(1)}
\\ & & \\ & & 
-180 (\alfa) DX^{\mu} DX^{{\nu}} B_{{\mu} {\nu}} 
(\partial\omega^{(1)})^2
\\ & & \\ & & 
+\frac{20}{2\pi\alpha^\prime}C^{(3)}(i_k B)(i_k C^{(3)}+
m\pi\alpha^\prime\omega^{(0)} B)\\ & & \\ & &
-\frac{15}{4}m^2(2\pi\alpha^\prime)
(\omega^{(0)})^2 DX^{\mu_1}\dots DX^{\mu_6}B^3_{\mu_1\dots\mu_6}
\\ & & \\ & &
-45m(2\pi\alpha^\prime)\omega^{(0)}DX^{\mu_1}\dots DX^{\mu_4}
B^2_{\mu_1\dots\mu_4}\partial\omega^{(1)}\\
& & \\ & &
+\frac{15}{2}m\omega^{(0)}B^2(i_k C^{(3)})
-360 (\alfa)^2 A (\partial \omega^{(1)}+
\frac{m}{4}\omega^{(0)}B)^2
\partial \omega^{(0)}
\\ & & \\ & & 
+ 15 (\alfa)^2 { e^{2 \phi} |k|^{-2} \left(i_k C^{(1)}+
m\pi\alpha^\prime\omega^{(0)}\right) \over 
1 + e^{2 \phi} |k|^{-2} 
\left(i_k C^{(1)}+m\pi\alpha^\prime\omega^{(0)}\right)^2}
{\cal K}^{(2)}{\cal K}^{(2)}{\cal K}^{(2)}
 \, .\\
\end{array}
\end{equation}

In Table \ref{table-mIIAKK} we summarize the worldvolume field content.
The gauge transformation rules of background and worldvolume fields
can be found in Appendices B, C.2 and reference \cite{BLO}.
We summarize our notation for the Type IIA background fields in
Table \ref{IIAback}.

\begin{table}[h]
\renewcommand{\arraystretch}{1.5}
\begin{center}
\begin{tabular}{|c|c|}
\hline
Worldvolume        & Field  \\
Field              & Strength  \\
\hline\hline
$ {\omega}^{(0)} $ &
${\cal K}^{(1)}_i$ \\
\hline
$ {\omega}^{(1)}_i$ &
 ${\cal K}^{(2)}_{ij}$ \\
\hline
$d^{(5)}$ & $ $  \\
\hline
${\omega}^{(5)}_{i_1\dots i_5}$
& ${\cal K}^{(6)}_{i_1\dots i_6}$ \\
\hline
${\omega}^{(6)}_{i_1\dots i_6}$
& \\
\hline
\end{tabular}
\end{center}
\caption{\label{table-mIIAKK} \small {\bf Worldvolume fields of the
massive IIA KK-monopole.}
In this table we give the
worldvolume fields, together
with their field strengths,
that occur in the effective
action of the massive IIA KK-monopole.}
\renewcommand{\arraystretch}{1}
\end{table}

\begin{table}[!ht]
\renewcommand{\arraystretch}{1.5}
\begin{center}
\begin{tabular}{|c|c|c|c|}
\hline
Target space & Gauge     & Dual  & Gauge    \\
Field        & Parameter & Field & Parameter \\
\hline\hline
${g}_{{\mu}{\nu}}$, $\phi$
& $-$ & $-$ & $-$ \\
\hline
$B_{\mu \nu}$ & $\Lambda_\mu$ & ${\tilde B}_{\mu_1 \dots \mu_6}$ 
& ${\tilde \Lambda}_{\mu_1 \dots \mu_5}$ \\
\hline
$C^{(1)}_\mu$ & $\Lambda^{(0)}$ & $C^{(7)}_{\mu_1 \dots \mu_7}$ 
& $\Lambda^{(6)}_{\mu_1 \dots \mu_6}$ \\ 
\hline
${C}^{(3)}_{{\mu}{\nu}{\rho}}$ & $\Lambda^{(2)}_{\mu \nu}$ & 
$C^{(5)}_{\mu_1 \dots \mu_5}$ 
& $\Lambda^{(4)}_{\mu_1 \dots \mu_4}$ \\
\hline
$k_{\mu}$ & $-$ & ${N}_{\mu_1 \dots \mu_7}$ & 
${\Omega}^{(6)}_{\mu_1\dots\mu_6}$ \\
\hline
\end{tabular}
\end{center}  
\caption{\label{IIAback} \small {\bf Target space fields of the type IIA 
superstring.}
The type IIA background contains the NS-NS sector:
$( g_{\mu \nu}, \phi, B_{\mu \nu})$,
the RR sector:
$( C^{(1)}, C^{(3)})$, 
and the Poincar{\'e} duals of the RR fields and the NS-NS 2-form $B$:
$( C^{(5)}, C^{(7)}, {\tilde B})$. The Kaluza-Klein monopole couples to a
new field $N$, dual to the Killing vector associated to the Taub-NUT
isometry, considered as a 1-form $k_\mu$.}
\renewcommand{\arraystretch}{1.5}
\end{table}

The action of the ten dimensional IIA KK-monopole is manifestly invariant
under translations of the Taub-NUT coordinate, since all the reduced fields
and gauge parameters have vanishing Lie derivative with respect to $k$. This 
symmetry has been restored in the reduction by a mechanism in which the
$y$ component of the Killing vector in eleven dimensions gives rise to an
auxiliary worldvolume field that is needed to compensate the massive
variation of the Stueckelberg field $i_k C^{(1)}$.  
A further check of this action is that for $m=0$ reduces to 
the action of the massless IIA KK-monopole \cite{BEL}.
 
It is also worth noting that the field $b$ has 
disappeared in the reduced action. This reflects the fact  that there 
are no fundamental strings
ending on the monopole. Nevertheless there is a string-like object,
described by $\omega^{(1)}$, ending on the monopole.
In fact, the worldvolume fields that couple
to the soliton solutions of a KK-monopole are those necessary to
construct invariant field strengths for the fields
$i_k C^{(p+1)}$ (see \cite{EJL}). 
These field strengths have the form:

\begin{equation}
{\cal K}^{(p)} = p\partial\omega^{(p-1)}+\frac{1}{2\pi\alpha^\prime}
(i_k C^{(p+1)})+\dots  \, ,
\end{equation}

\noindent so that $\omega^{(p-1)}$ couples to a
$(p-2)$-brane soliton which describes the boundary of a $p$-brane
ending on the monopole, with one of its worldvolume directions
wrapped around the Taub-NUT direction of the monopole. Since the
target space field associated to $\omega^{(1)}$ is $(i_k C^{(3)})$
it describes a wrapped D2-brane ending on the monopole.

We also find the soliton configurations: 
$(2|{\rm D}4,{\rm KK})$, $(3|{\rm NS}5,KK)$ and $(3|{\rm KK},{\rm KK})_{1,2}$ 
(see \cite{Papa,EJL}). The only modifications due to the mass occur
in the explicit expressions of the field strengths, where 
new terms proportional to the mass appear, which involve
worldvolume fields that already propagated in the massless case.
There is however an exception, and that is the presence of the 
$d^{(5)}$ worldvolume field associated to the dual massive
transformations, which couples to a 4-brane soliton.
This soliton is a domain wall in the six dimensional
worldvolume and is described by the configuration:

\begin{displaymath}
 (4|{\rm D}8,{\rm KK})=
\mbox{ 
$\left\{ \begin{array}{c|ccccccccc}
         \x & \x & \x  & \x & \x & - & \x & \x & \x  & \x \\
         \x & \x & \x   & \x  & \x & \x & z & - & -  & -           
                                    \end{array} \right.$}   
\end{displaymath}

\noindent which is obtained by reducing the $(4|{\rm M}9,{\rm KK})$ soliton
configuration of the M-theory KK-monopole along the isometric
direction of the M9-brane.
This intersection is related to the Hanany-Witten 
configuration \cite{HW}:

\begin{equation}
\label{conf1}
 \begin{array}{c|c}
{\rm D5}:\ \ \          \x & \x    \x   -   -   -  \x  \x   \x   - \\
{\rm NS5}:\          \x & \x   \x  \x  \x  \x  -  -   -   -    \\
{\rm D3}: \ \ \       \x & \x \x - - - - - - \x  
                         \end{array} 
\end{equation}

\noindent by T-duality along the 3,4,9 directions.
It is also T-dual to the intersection $(4|{\rm D}7,{\rm NS}5)$ in Type IIB,
corresponding to a 4-brane soliton in the IIB NS-5-brane
\cite{EGJP}.
 
There is as well another 4-brane soliton in the KK-monopole
worldvolume \cite{BREJS} which is already present in the
massless case. This is the embedding of the D4-brane on
the monopole: $(4|{\rm D}4,{\rm KK})$, and it couples to the worldvolume
field $\omega^{(5)}$, describing the tension of the monopole.
This is the reduction of the 5-brane soliton $(5|{\rm M}5,{\rm KK})$, which
couples to the tension ${\hat \omega}^{(6)}$ of the M-KK-monopole.
Finally, the reduction of the 6-brane soliton $(6|{\rm M}9,{\rm KK})$ 
gives a 5-brane
soliton realized as the embedding of the KK-monopole on a
KK-7A-brane\footnote{This brane is obtained by reducing the M9-brane
along a worldvolume direction other than the $z$-direction,
but it is not predicted by the Type IIA supersymmetry algebra.
As discussed in \cite{proci}, this is also the case for the 
KK-6A brane, obtained by reducing the M-KK-monopole along a
transverse direction different from the Taub-NUT direction.
These branes are required by U-duality of M-theory on a 
d-torus, as it has been shown in \cite{U1,U2,U3}.}.


\section{Massive $T$-duality:\\
 IIB NS-$5$ $\rightarrow$ Massive IIAKK}
\label{IIBNS5-->IIAKK}
         

In this section we obtain the massive IIA KK-monopole through
a ``massive'' T-duality transformation in the action of the
IIB NS-5-brane.

The action of the Type IIB NS-5-brane was constructed in 
\cite{EJL} and is given by:

\begin{eqnarray}
\label{5brane-action}
\hspace{-.5cm} S &=&
-T_{{\rm NS-5B}} \int d^6 \xi \,\,
e^{-2\varphi} \sqrt{1 + e^{2 \varphi} (C^{(0)})^2} \, \times
\nonumber \\
&&\hspace{+1cm}
\times \, \sqrt{|{\rm det} 
( {\j} - (\alfa) {e^{\varphi} \over \sqrt{1 + e^{2 \varphi}
(C^{(0)})^2}} {\tilde {\cal F}} )|} \, \nonumber \\
&&\hspace{-.5cm}+ \,\, {1 \over 6!}(\alfa) T_{{\rm NS-5B}} \int d^6\xi
\,\,  \epsilon^{i_1 \dots i_6}
{\tilde {\cal G}}^{(6)}_{i_1 \dots i_6} \, .
\end{eqnarray}
Here 
${\tilde {\cal F}}=2\partial c^{(1)}+\frac{1}{2\pi\alpha^\prime} C^{(2)}$
and:
\begin{equation}
\begin{array}{rcl}
{\tilde {\cal G}}^{(6)} &=& \left\{
6\partial {\tilde c}^{(5)} - {1 \over \alfa}{\tilde {\cal B}}
- {45 \over 2(\alfa)} {\cal B}C^{(2)}C^{(2)} -15 C^{(4)} {\tilde {\cal F}} 
\right. \\ & &\\
& & - 180 (\alfa) {\cal B} \partial c^{(1)}\partial c^{(1)}
-90 {\cal B} C^{(2)} \partial c^{(1)}
\\& &\\& &
\left.
+15 (\alfa)^2 {C^{(0)} \over e^{-2 \varphi} + (C^{(0)})^2} 
{\tilde {\cal F}}{\tilde {\cal F}}{\tilde {\cal F}}
\right\} \, .\\
\end{array}
\end{equation}

In Table \ref{IIBback} we have summarized our notation for the 
Type IIB background fields.

\begin{table}[!ht]
\renewcommand{\arraystretch}{1.5}
\begin{center}
\begin{tabular}{|c|c|c|c|}
\hline
Target space & Gauge     & Dual  & Gauge    \\
Field        & Parameter & Field & Parameter \\
\hline\hline
${\j}_{{\mu}{\nu}}$, $\varphi$
& $-$ & $-$ & $-$ \\
\hline
${\cal B}_{\mu \nu}$ & $\Lambda_\mu$ & ${\tilde {\cal B}}_{\mu_1 \dots \mu_6}$
 & ${\tilde \Lambda}_{\mu_1 \dots \mu_5}$ \\
\hline
$C^{(0)}$ & $-$ & $-$ & $-$ \\ 
\hline
${C}^{(2)}_{{\mu}{\nu}}$ & $\Lambda^{(1)}_\mu$ 
& $C^{(6)}_{\mu_1 \dots \mu_6}$ 
& $\Lambda^{(5)}_{\mu_1 \dots \mu_5}$ \\
\hline
\end{tabular}
\end{center}  
\caption{\label{IIBback} \small 
{\bf Target space fields of the type IIB superstring.}
The Type IIB background contains 
the common sector:
$( {\j}_{\mu \nu}, \varphi, {\cal B}_{\mu \nu} )$,
the RR sector: $(C^{(0)}, C^{(2)}, C^{(4)})$, 
and the Poincar{\'e} duals of the 2-forms $C^{(2)}$ and ${\cal B}$:  
$( C^{(6)}, {\tilde {\cal B}})$. 
}
\renewcommand{\arraystretch}{1}
\end{table}

We apply now a T-duality transformation along a transverse
direction. The worldvolume fields do not change rank after
T-duality since the original and dual branes have the same
number of worldvolume dimensions.
Moreover, we have a new scalar field $Z'$, which is the T-dual
of the coordinate along which we perform the duality transformation.
The T-duality rules for the worldvolume fields are given by:
 
\begin{equation}
\begin{array}{rcl}
Z' &=& (\alfa) \omega^{(0)} \, ,\\
& &\\
c^{(1) \prime} &=& - \omega^{(1)} 
- {m \over 4} (\alfa) (\omega^{(0)})^2 \partial Z \, ,\\
& &\\
\partial{\tilde c}^{(5) \prime} &=& \partial\omega^{(5)} +
60 (\alfa)^2 \partial Z \partial \omega^{(1)} \partial \omega^{(1)}
\partial \omega^{(0)}
\\& &\\
& &+m\omega^{(6)}-\frac12 m(2\pi\alpha^\prime)
d^{(5)}\partial\omega^{(0)}\, .
\end{array}
\end{equation}

\noindent Here $Z$ is the Taub-NUT coordinate of the
KK-monopole. Its occurrence on the right hand side of the two
expressions above is required by gauge invariance, and assures
the gauged sigma-model structure necessary to describe the
KK-monopole.

Using the massive T-duality rules given in (\ref{IIB-IIA-massive}) and
(\ref{atres}) for the background fields, 
we find the following transformations for the worldvolume
curvatures:

\begin{equation}
{\tilde {\cal F}}^{\prime} = - {\cal K}^{(2)} \, ,\qquad
{\tilde {\cal G}}^{(6) \prime}= {\cal K}^{(6)} \, ,
\end{equation}

\noindent where ${\cal K}^{(2)}$ is given by
(\ref{curvas}) and ${\cal K}^{(6)}$ by (\ref{WZKK}). 
Substituting in the IIB NS-5-brane effective action we 
recover the expression (\ref{accionmasiva})
for the effective action of the massive IIA KK-monopole that we
obtained from eleven dimensions.
This provides a check of that action as well as of the massive
T-duality rule (\ref{atres}) given in the Appendix.

One remark is in order at this point. In general the double 
dimensional reduction of a massive M-brane gives a tension
\cite{BLO}: 

\begin{equation}
T=\left( 1- {m \over 2} (\alfa) v^{(0)} \right) {\hat T} \, \int d {\hat \xi}
\end{equation}
 
\noindent for the reduced brane, where ${\hat \xi}$ is the
compact worldvolume direction. A particular example is the massive
IIA KK-monopole obtained in the previous section. 
This suggests that in the massive case the tensions should transform 
under T-duality as:

\begin{equation}
T^\prime_{{\rm B5}}=T_{{\rm mAKK}}=
\left( 1- {m \over 2} (\alfa) v^{(0)} \right) T_{{\rm mMKK}}\, ,
\end{equation}

\noindent which seems, however, an arbitrary choice from the 
T-duality point of view. We leave some especulations about this point for the
Conclusions.


\section{Conclusions}


We have constructed the worldvolume effective action of the 
Type IIA KK-monopole propagating in a background with 
non-vanishing cosmological constant. The worldvolume field
content of this brane consists on a scalar, a 1-form, two
5-forms and a 6-form. One of the two 5-forms is interpreted as
the tension of the monopole, whereas the other one is 
associated to dual massive transformations in the worldvolume
of the monopole. The 6-form and the tension of the monopole 
are interpreted as a Stueckelberg pair with respect to massive
transformations. 

These worldvolume fields have been interpreted
in terms of soliton solutions propagating in the worldvolume of
the monopole. The new feature with respect to the massless case
is that there is a 4-brane soliton realized as the domain wall
intersection of the KK-monopole with a D8-brane, and a 6-brane
soliton corresponding to the embedding of the monopole on
a KK7-brane. 

We have already mentioned that the KK7-brane belongs
to a particular class of Type IIA brane solutions which are not
predicted by the spacetime supersymmetry algebra. It is unclear
why this happens. These branes have been encountered already
in the literature in a different context, namely they are required
in order to fill up multiplets of BPS states in representations
of the U-duality symmetry group of M-theory on a d-torus
\cite{U1,U2,U3,MO}. As we discuss in \cite{proci}, where we
calculate the kinetic part of their worldvolume effective actions,
these branes do not have an obvious interpretation in weakly coupled 
string theory  since they scale with the coupling constant more singularly
than $1/{g_s^2}$.

The derivation of the KK-monopole effective action from eleven 
dimensions gives  a tension depending on the mass and on a Wilson line
of the gauge field associated to the massive isometry
(which is set to a constant value
by the integration of the worldvolume 5-form playing the role
of tension of the monopole) (see (\ref{modifi})).
In particular, the tension can vanish for a certain value
of the Wilson line. In fact, this is a general feature for the Type IIA
massive branes that are obtained from eleven dimensions by a
double dimensional reduction \cite{BLO}.
The presence of covariant derivatives with respect to massive
transformations in the eleven-dimensional massive branes implies
that in the double dimensional reduction the brane can be 
wrapped along the eleventh direction with unusual
winding number, which depends on the mass and on a Wilson line
of the gauge field. 
Therefore, consistency implies that $(1-\frac{m}{2}(2\pi\alpha^\prime)
v^{(0)})$ has to be an integer, i.e. the winding number, with
the tension of the reduced brane proportional to it.
The Wilson line may be interpreted as a distance 
$\alfa v^{(0)}$ corresponding to a D8-brane, 
since a massive IIA brane is a brane in the presence
of D8-branes. 
This is similar to the interpretation of the masses
of the string states in terms of the D8-brane positions 
in a Type ${\rm I}^\prime$ construction.
Since the D8-branes are T-dual to the
D9-branes, the arbitrariness of $v^{(0)}$ from the T-duality point
of view might be originated by Wilson lines in Type IIB with
D9-branes.

Finally it would be of interest to analyze the six dimensional
interacting gauge theory that arises from considering a system
of parallel massive KK-monopoles in the limit in which gravity is
decoupled \cite{sixdi}.

\section*{Acknowledgements}

We would like to thank E.~Bergshoeff for useful discussions.
The work of E.E.~ is part of the research program of the
Dutch Foundation FOM.

\appendix


\section{$T$-duality}
\label{T-duality}


In this Appendix we summarize the massive T-duality rules for the
background fields given in \cite{BRGPT, Eric-Mees,
 Green-Hull-Townsend} \footnote{The T-duality rules between Type IIA
and Type IIB backgrounds were derived in \cite{BHO} for the
massless case.}.
We also derive the massive T-duality transformation of the
${\tilde {\cal B}}$ field coupled to the IIB NS-5-brane. 
In our notation 
$z$ is the direction along which we perform the duality transformation.

The following T-duality rules from Type IIB\footnote{In the basis
in which $C^{(4)}$ is S-duality invariant.}
onto massive Type IIA
backgrounds have already been constructed in the literature:

\begin{equation}
\begin{array}{rcl}
C^{(0) \prime} &=& - C^{(1)}_z - {m \over 2}{\tilde Z} \, ,\\
& &\\
C^{(2) \prime}_{\mu z} &=& -C^{(1)}_\mu + 
\left(C^{(1)}_z  + {m \over 2} {\tilde Z}
 \right){ g_{\mu z} \over g_{zz}}\, ,\\
& &\\
C^{(2) \prime}_{\mu \nu} &=& - C^{(3)}_{\mu \nu z} + 2C^{(1)}_{[\mu} B_{\nu] z}
\\& &\\& &- {2}C^{(1)}_z {g_{z [\mu} \over g_{zz}}B_{\nu] z} \\
& &\\
& &- {m \over 2} {\tilde Z} \left( B_{\mu \nu} - 2 B_{[\mu z} {g_{\nu]z} \over g_{zz}}
\right) \, ,\\
& &\\
C^{(4) \prime}_{\mu \nu \rho z} &=& - C^{(3)}_{\mu \nu \rho}
+{3 \over 2} C^{(3)}_{[\mu \nu z}{g_{\rho] z} \over g_{zz}} 
\\& &\\& &
+ {3 \over 2} \left( C^{(1)}_{[\mu}-C^{(1)}_z {g_{[\mu z} \over g_{zz}}
\right) B_{\nu \rho]}\, ,\\
\end{array}
\begin{array}{rcl}
e^{\varphi^{\prime}} &=& {1 \over \sqrt{|g_{zz}|}} e^{\phi} \, ,\\
& &\\
{\cal B}^{\prime}_{\mu \nu} &=&
 B_{\mu \nu} - {2 \over g_{zz}} B_{[\mu z}g_{\nu] z} 
\, ,\\ & &\\
{\cal B}^{\prime}_{\mu z} &=& -{g_{\mu z} \over g_{zz}} \, ,\\
& &\\
\j^{\prime}_{\mu \nu} &=& g_{\mu \nu} - {1 \over g_{zz}} \left(
g_{\mu z} g_{\nu z} - B_{\mu z} B_{\nu z} \right) \, ,\\
& &\\
\j^{\prime}_{\mu z} &=&  - {1 \over g_{zz}}B_{\mu z} \, ,\\
& &\\
\j^{\prime}_{zz} &=& {1 \over g_{zz}} \, ,\\
\end{array}
\label{IIB-IIA-massive}
\end{equation}
\begin{displaymath}
\begin{array}{rcl}
C^{(4) \prime}_{\mu_1 \dots \mu_4} &=&
-C^{(5)}_{\mu_1 \dots \mu_4 z}
+4 \left( C^{(3)}_{[\mu_1 \mu_2 \mu_3}
-3 C^{(3)}_{[\mu_1 \mu_2 z} {g_{\mu_3 z} \over g_{zz}} \right)B_{\mu_4] z} 
\\& &\\& &
+3 C^{(3)}_{[\mu_1 \mu_2 z} \left(
B_{\mu_3 \mu_4]} -2 B_{\mu_3 z} {g_{\mu_4] z} \over g_{zz}} \right)
-6 \left( C^{(1)}_{[\mu_1} - C^{(1)}_z {g_{[\mu_1 z} \over g_{zz}}
\right) B_{\mu_2 \mu_3} B_{\mu_4] z}\, ,\\
& &\\
C^{(6) \prime}_{\mu_1 \dots \mu_5 z} &=&
-C^{(5)}_{\mu_1 \dots \mu_5}
+5 C^{(5)}_{[\mu_1 \dots \mu_4 z}{g_{\mu_5] z} \over g_{zz}}
\\& &\\& &
-15 C^{(3)}_{[\mu_1 \mu_2 z} B_{\mu_3 \mu_4} {g_{\mu_5] z} \over g_{zz}}
\\& &\\& &
+ {15 \over 2} \left(
C^{(1)}_{[\mu_1} - C^{(1)}_z {g_{[\mu_1 z} \over g_{zz}}
- {m \over 2} {\tilde Z} {g_{[\mu_1 z} \over g_{zz}} \right)
B_{\mu_2 \mu_3}B_{\mu_4 \mu_5]} \, .\\
\end{array}
\end{displaymath}

The inverse rules (from massive Type IIA to Type IIB) are given by:

\begin{displaymath}
\begin{array}{rcl}
C^{(1) \prime}_z &=& - C^{(0)} - {m\over 2}{\tilde Z} \, ,\\
& &\\
C^{(1) \prime}_\mu &=& - C^{(2)}_{\mu z} + C^{(0)} {\cal B}_{\mu z} \, ,\\
& &\\
C^{(3) \prime}_{\mu \nu z} &=& - C^{(2)}_{\mu \nu} +
{2} C^{(2)}_{[\mu z} {\j_{\nu ] z} \over \j_{zz}}\\
& &\\
& &- {m \over 2} {\tilde Z} \left( {\cal B}_{\mu \nu} 
- 2 {\cal B}_{[\mu z} { \j_{\nu] z} \over \j_{zz}} \right) \, ,\\
& &\\
C^{(3) \prime}_{\mu \nu \rho} &=& - C^{(4)}_{\mu \nu \rho z}
+{3 \over 2}C^{(2)}_{[\mu \nu}{\cal B}_{\rho] z}
\\& &\\& &
-{3 \over 2}{\cal B}_{[\mu \nu}C^{(2)}_{\rho] z}
\\& &\\& &
+ 6 C^{(2)}_{[\mu z}{\cal B}_{\nu z} {\j_{\rho] z} \over \j_{zz}}\, ,\\
\end{array}
\begin{array}{rcl}
e^{\phi^{\prime}} &=& {1 \over \sqrt{|\j_{zz}|}}  e^{\varphi} \, ,\\
& &\\
g^{\prime}_{\mu \nu} &=& \j_{\mu \nu} - {1 \over \j_{zz}}
\left(\j_{\mu z} \j_{\nu z} - {\cal B}_{\mu z} {\cal B}_{\nu z} \right)
\, ,\\
& &\\
g^{\prime}_{\mu z} &=& - {1 \over \j_{zz}} {\cal B}_{\mu z} \, ,\\
& & \\
g^{\prime}_{zz} &=& {1 \over \j_{zz}} \, ,\\
& & \\
B^{\prime}_{\mu \nu} &=& {\cal B}_{\mu \nu} 
- {2 \over {\j}_{zz}}{\cal B}_{[\mu z} \j_{\nu] z} \, ,\\
& &\\
B^{\prime}_{\mu z} &=& - { \j_{z \mu} \over \j_{zz}} \, ,\\
\end{array}
\end{displaymath}
\begin{equation}
\begin{array}{rcl}
C^{(5) \prime}_{\mu_1 \dots \mu_4 z} &=& - C^{(4)}_{\mu_1 \dots \mu_4}
+ 4  C^{(4)}_{[\mu_1 \mu_2 \mu_3 z} {\j_{\mu_4] z} \over \j_{zz}}
- 3 C^{(2)}_{[\mu_1 \mu_2} {\cal B}_{\mu_3 \mu_4]}
\\& &\\& &
-6C^{(2)}_{z [\mu_1} {\cal B}_{\mu_2 \mu_3} {\j_{\mu_4] z} \over \j_{zz}}
-6{\cal B}_{z [\mu_1}C^{(2)}_{\mu_2 \mu_3} {\j_{\mu_4] z} \over \j_{zz}} 
\\& &\\& &-{3 \over 2}m {\tilde Z}
 {\cal B}_{[\mu_1 \mu_2}{\cal B}_{\mu_3 \mu_4]}
+ 6 m {\tilde Z} {\cal B}_{[\mu_1 \mu_2} {\cal B}_{\mu_3 z}{\j_{\mu_4]z} \over \j_{zz}}
\, ,\\
& &\\
C^{(5) \prime}_{\mu_1 \dots \mu_5} &=& -C^{(6)}_{\mu_1 \dots \mu_5 z}
+5 \left( C^{(4)}_{[\mu_1 \dots \mu_4}
- 4 C^{(4)}_{[\mu_1 \dots \mu_3 z} 
{\j_{\mu_4 z} \over \j_{zz}}\right){\cal B}_{\mu_5]z}
\\& &\\& &
-{15 \over 2}{\cal B}_{[\mu_1 \mu_2}
{\cal B}_{\mu_3 \mu_4]}C^{(2)}_{\mu_5] z} 
- 30 C^{(2)}_{z [\mu_1} {\cal B}_{\mu_2 \mu_3}{\cal B}_{\mu_4 z}
{\j_{\mu_5] z} \over \j_{zz}}\, .\\ 
\end{array}
\end{equation}

The IIB NS-5-brane couples minimally to the field ${\tilde {\cal B}}$,
dual to the NS-NS 2-form. Its massive T-duality rules are given by:
 
\begin{equation}
\label{atres}
\begin{array}{rcl}
{\tilde {\cal B}}^{\prime}_{\mu_1 \dots \mu_5 z} &=&
{\tilde B}_{\mu_1 \dots \mu_5 z} 
-  5 \left( C^{(5)}_{[\mu_1 \dots \mu_4 z}
-3B_{[\mu_1 \mu_2} C^{(3)}_{\mu_3 \mu_4 z}\right)
\left(C^{(1)}_{\mu_5]} - C^{(1)}_z {g_{\mu_5] z} \over g_{zz}} \right) \\
& &\\& & 
-5 \left(C^{(3)}_{[\mu_1 \mu_2 \mu_3} 
- {3 \over 2} C^{(3)}_{[\mu_1 \mu_2 z}
{g_{\mu_3] z} \over g_{zz}}\right)
C^{(3)}_{\mu_4 \mu_5 z}
\\& &\\& &
- {m \over 2} {\tilde Z} \left( C^{(5)}_{\mu_1 \dots \mu_5}
-5 C^{(5)}_{[\mu_1 \dots \mu_4 z} {g_{\mu_5] z} \over g_{zz}} \right)
\\& &\\& &
+ {15 \over 8} m^2 {\tilde Z}^2 B_{[\mu_1 \mu_2}B_{\mu_3 \mu_4}
{g_{\mu_5] z} \over g_{zz}} 
\, ,
\\
& &
\\
{\tilde {\cal B}}^{\prime}_{\mu_1 \dots \mu_6} &=& 
- N_{\mu_1 \dots \mu_6 z} 
-6 {\tilde B}_{[\mu_1 \dots \mu_5 z}B_{\mu_6] z} 
\\& &\\& &
+30\left(C^{(5)}_{[\mu_1 \dots \mu_4 z} 
-3 C^{(3)}_{[\mu_1 \mu_2 z}B_{\mu_3 \mu_4}\right)
\left(C^{(1)}_{\mu_5} - C^{(1)}_z 
{g_{\mu_5 z} \over g_{zz}} \right) B_{\mu_6] z}\\
& &\\& &
+10 \left(C^{(3)}_{[\mu_1 \dots \mu_3} 
-{3 \over 2} {g_{[\mu_1 z} \over g_{zz}} C^{(3)}_{\mu_2 \mu_3 z} \right)
\left(C^{(3)}_{\mu_4 \mu_5 z} B_{\mu_6] z}
- m {\tilde Z} B_{\mu_4 \mu_5}B_{\mu_6] z} \right)
\\& &\\& &
+3m {\tilde Z} \left( C^{(5)}_{[\mu_1 \dots \mu_5}
-5C^{(5)}_{\mu_1 \dots \mu_4 z}{g_{\mu_5 z} \over g_{zz}}\right)
B_{\mu_6] z}
\\& &\\& &
-30 C^{(3)}_{[\mu_1 \mu_2 z}
\left(C^{(3)}_{\mu_3 \mu_4 z} + {m \over 2}{\tilde Z} B_{\mu_3 \mu_4}\right)
{g_{\mu_5 z} \over g_{zz}}B_{\mu_6] z}
\\& &\\& &
- {15 \over 2} \left(C^{(3)}_{[\mu_1 \mu_2 z}
+ {m \over 2} {\tilde Z} B_{[\mu_1 \mu_2} \right)
\left(C^{(3)}_{\mu_3 \mu_4 z} + {m \over 2} {\tilde Z} B_{\mu_3 \mu_4}
\right) B_{\mu_5 \mu_6}
\\& &\\& &
+ {45 \over 4} m^2 {\tilde Z}^2 B_{[\mu_1 \mu_2}
B_{\mu_3 \mu_4}B_{\mu_5 z} {g_{\mu_6] z} \over g_{zz}}
\, .\\
\end{array}
\end{equation}

Let us clarify the role played by ${\tilde Z}$ in the expressions
above. In the $T$-duality from the IIB (IIA) NS-$5$-brane to the 
IIA (IIB) KK-monopole,
${\tilde Z}$ is the T-dual transformed of the coordinate transversal
to the NS-5-brane:

\begin{equation}
{\tilde Z}=Z'=(2\pi\alpha^\prime)\omega^{(0)}\, ,
\end{equation}

\noindent and not another embedding coordinate. However, in the
T-duality from the IIA (IIB) KK-monopole to the IIB (IIA) 
NS-5-brane, ${\tilde Z}$ is indeed the transversal coordinate
to the 5-brane. When the T-duality takes place
between two D-branes ${\tilde Z}$ is proportional to the 
$\sigma$ component of the Born-Infeld field of
the dual brane when the reduction is direct, or is equal to 
$\sigma$ when the reduction is 
double\footnote{The double dimensional reduction of a
Type IIB brane must be a Scherk-Schwarz type of reduction 
in order to make contact with massive Type IIA.}.


\section{Target Space Gauge Symmetries}


The gauge symmetries of the
eleven-dimensional background fields that couple to the
M-KK-monopole are given by:

\begin{eqnarray}
\label{num}
\delta {\hat g}_{{\hat \mu} {\hat \nu}} &=&
 - m (i_{\hat h}{\hat \chi})_{({\hat \mu}}
(i_{\hat h} {\hat g})_{{\hat \nu})}
\, ,\nonumber \\
\delta {\hat C}&=& 3\partial{\hat \chi}
 - {3 \over 2}m (i_{\hat h}{\hat \chi})(i_{\hat h}{\hat C})\, ,
\nonumber \\
\delta \ikC &=&
2\partial (i_{{\hat k}}{\hat \chi})
-m(i_{\hat h}{\hat \chi})(i_{\hat k}i_{\hat h}
{\hat C})-m(2\pi\alpha^\prime)\partial {\hat \omega}^{(0)}
(i_{\hat h}{\hat \chi})
\, ,
\nonumber \\
\delta \iktildeC
  &=& 5\partial (i_{{\hat k}}{\hat {\tilde \chi}})
+15 \partial{\hat \chi}(i_{\hat k}{\hat C}) 
-10{\hat C}\partial(i_{\hat k}{\hat \chi})\\
& & +\frac52 m (i_{\hat h}{\hat \chi})
(i_{\hat k}i_{\hat h}{\hat {\tilde C}})
+\frac{m}{2} (i_{\hat h}i_{\hat k}{\hat \Sigma})
-\frac52 m (2\pi\alpha^\prime)\partial {\hat \omega}^{(0)}
(i_{\hat h}{\hat {\tilde \chi}})
\, , \nonumber \\
\delta (i_{\hat k}{\hat N}) &=&
7 \left\{ \partial (i_{\hat k} {\hat  \Omega}) 
+15 \partial\iktildechi \ikC  
+30 \partial {\hat \chi}{\ikC}{\ikC}
 \right.\nonumber \\
& & \left.
 -20 {\hat C}\ikC \partial \ikchi +\frac{m}{2} (i_{\hat h}
{\hat \chi})(i_{\hat k}i_{\hat h}{\hat N})
-\frac{m}{2}(2\pi\alpha^\prime)\partial {\hat \omega}^{(0)}
(i_{\hat h}{\hat \Sigma})\right. \nonumber \\
& & \left.
+\frac32 m (i_{\hat h}i_{\hat k}{\hat \Sigma})
(i_{\hat k}{\hat C})\right. \nonumber \\
& & \left.
-\frac{15}{2}m(2\pi\alpha^\prime)
\partial {\hat \omega}^{(0)}(i_{\hat h}{\hat {\tilde \chi}})
(i_{\hat k}{\hat C})+10m(2\pi\alpha^\prime){\hat C}
(i_{\hat k}{\hat C})\partial {\hat \omega}^{(0)}
(i_{\hat h}{\hat \chi})
\right\} \, .
\end{eqnarray}

\noindent These fields transform as well under ${\hat \rho}^{(0)}$ and
${\hat \sigma}^{(0)}$ transformations in the usual way.

The transformations of the ten dimensional fields coupled to
the massive IIA KK-monopole can be found in \cite{BLO}, as
well as the relations between ten- and eleven- dimensional fields 
and gauge parameters. The transformation of $i_k N$ is however
new, and it is given by:

\begin{equation}
\begin{array}{rcl}
\label{num2}
\delta (i_k N) &=&
6 \partial (i_k \Omega^{(6)})-3m(i_k \Sigma^{(6)})(i_k B)\\
& &\\
& &
+60 \partial (i_k \Lambda^{(4)})(i_k C^{(3)}+\frac{m}{2}
(2\pi\alpha^\prime)\omega^{(0)}B)\\ 
& &\\
& & 
 -30 (\partial (i_k {\tilde \Lambda})-\frac{m}{2}
(2\pi\alpha^\prime)\omega^{(0)}\partial\Lambda^{(4)})
(i_k B)\\
& &\\
& &
-60\partial\Lambda (i_k C^{(3)}+\frac{m}{2}
(2\pi\alpha^\prime)\omega^{(0)}B)^2\\
& &\\
& &
+120 \partial\Lambda^{(2)}(i_k C^{(3)}+\frac{m}{2}
(2\pi\alpha^\prime)\omega^{(0)}B)(i_k B)\\
& &\\
& &
+60 (\partial (i_k \Lambda^{(2)})+\frac{m}{2}
(2\pi\alpha^\prime)\omega^{(0)}\partial\Lambda)
B(i_k C^{(3)}+\frac{m}{2}(2\pi\alpha^\prime)\omega^{(0)}B)\\
& &\\
& &
+20\partial (i_k\Lambda)C^{(3)} (i_k C^{(3)}+\frac{m}{2}
(2\pi\alpha^\prime)\omega^{(0)}B)\\
& &\\
& &
-40 (\partial(i_k \Lambda^{(2)})+\frac{m}{2}
(2\pi\alpha^\prime)\omega^{(0)}\partial\Lambda) C^{(3)}
(i_k B) \, ,\\
\end{array}
\end{equation}

\noindent where: 

\begin{equation}
\begin{array}{rcl}
(i_{\hat k}{\hat \Omega})_{\mu_1\dots \mu_5 y}&=&
-(i_k \Omega^{(6)})_{\mu_1\dots\mu_5} \\
& & \\
(i_{\hat k}{\hat \Sigma})_{\mu_1\dots\mu_5 y}&=&
-(i_k \Sigma^{(6)})_{\mu_1\dots\mu_5}\, .\\
\end{array}
\end{equation}

\section{Worldvolume Gauge Symmetries}


Here we give the gauge transformations of the worldvolume fields 
present in the effective actions of the massive M-theory and Type IIA
KK-monopoles.


\subsection{Massive M-KK-monopole}
\label{massiveMKK}


\begin{equation}
\begin{array}{rcl}
\delta {\hat \omega}^{(1)}&=&\partial {\hat \mu}
-\frac{1}{2\pi\alpha^\prime}
(i_{\hat k}{\hat \chi})-\frac{m}{2}(2\pi\alpha^\prime)
\partial {\hat \omega}^{(0)}{\hat \rho}^{(0)}
 \, , \\
& & \\
\delta {\hat d}^{(5)}&=& \frac{1}{2\pi\alpha^\prime}
(i_{\hat h}i_{\hat k}{\hat \Sigma}) \, , \\
& & \\
\delta {\hat \omega}^{(6)} &=&
6 \partial{\hat \rho}^{(5)} -m {\hat \rho}^{(6)}
+ {1 \over \alfa} (i_{\hat k} {\hat \Omega})
- 30{\hat \omega}^{(1)} \partial
 (i_{\hat k} {\hat {\tilde \chi}})
\\ & & \\ & &
-180 (\alfa) \partial
 {\hat \chi} {\hat \omega}^{(1)}
 \partial {\hat \omega}^{(1)}
- 120 (\alfa)^2 \partial {\hat \sigma}^{(0)}
 {\hat \omega}^{(1)} \partial
 {\hat \omega}^{(1)} \partial
 {\hat \omega}^{(1)}  \, ,\\
& & \\
\delta {\hat \omega}^{(7)}
&=&\partial {\hat \rho}^{(6)}
+15 (2\pi\alpha^\prime)
(i_{\hat h}{\hat {\tilde \chi}})\partial {\hat \omega}^{(0)}
\partial {\hat \omega}^{(1)}-\frac12 \partial {\hat \omega}^{(0)}
(i_{\hat h}{\hat \Omega})
\\ & & \\ & &
-\frac32 (2\pi\alpha^\prime) {\hat d}^{(5)}
(\frac{2}{2\pi\alpha^\prime}\partial (i_{\hat k}{\hat \chi})
+m\partial (i_{\hat k}i_{\hat h}{\hat \chi}){\hat b}-
m\partial{\hat \omega}^{(0)} (i_{\hat h}{\hat \chi}))
\\ & & \\ & &
-3{\hat b}\left[ \partial (i_{\hat h}i_{\hat k}{\hat \Omega})-
5(2\pi\alpha^\prime)\partial (i_{\hat k}{\hat {\tilde \chi}})
\partial {\hat \omega}^{(0)}-20 (2\pi\alpha^\prime)\partial
(i_{\hat h}i_{\hat k}{\hat {\tilde \chi}})\partial 
{\hat \omega}^{(1)} \right.
\\ & & \\ & &
\left. +60 (2\pi\alpha^\prime)^3\partial {\hat \sigma}^{(0)}
(\partial {\hat \omega}^{(1)})^2\partial {\hat \omega}^{(0)}+
60(2\pi\alpha^\prime)^2\partial {\hat \chi}\partial 
{\hat \omega}^{(1)}\partial {\hat \omega}^{(0)} \right.
\\ & & \\ & &
\left. +60 (2\pi\alpha^\prime)^2\partial (i_{\hat h}{\hat \chi})
(\partial {\hat \omega}^{(1)})^2\right]  
\, . \\
\end{array}
\end{equation}


\subsection{Massive IIA KK-monopole}


\begin{equation}
\begin{array}{rcl}
\delta v^{(0)} &=& 0\, ,\\ 
& & \\
\delta \omega^{(0)} &=& \frac{1}{2\pi\alpha^\prime}
(i_k \Lambda) \, ,\\ 
& & \\
\delta {\omega}^{(1)}&=&\partial \mu^{(0)}-\frac{1}{2\pi\alpha^\prime}
(i_k \Lambda^{(2)})-\frac{m}{2}\omega^{(0)}\Lambda+
\omega^{(0)}(\partial\Lambda^{(0)}+ {m \over 2} (\alfa) 
\partial(\sigma^{(0)}\omega^{(0)}))
\, , \\ 
& & \\
\delta d^{(5)}&=&-\frac{1}{2\pi\alpha^\prime}(i_k \Sigma^{(6)})
\, ,\\ 
& & \\
\delta \omega^{(5)}&=&5\partial\rho^{(4)}-m\rho^{(5)}-
\frac{1}{2\pi\alpha^\prime}(i_k\Omega^{(6)})-5\partial 
(i_k {\tilde \Lambda})\omega^{(0)}+
20 \omega^{(1)}\partial (i_k\Lambda^{(4)})
\\ & & \\
&&-60(2\pi\alpha^\prime)\partial\Lambda^{(2)}\omega^{(1)}
\partial\omega^{(0)}
+60(2\pi\alpha^\prime)\omega^{(1)}
\partial\omega^{(1)}
\partial\Lambda \\ & & \\
&&+60(2\pi\alpha^\prime)^2\sigma^{(0)}
\partial\omega^{(1)}\partial \omega^{(1)} \partial\omega^{(0)} 
+\frac54 m (2\pi\alpha^\prime)(\omega^{(0)})^2\partial\Lambda^{(4)}
\, ,\\
& & \\
\delta \omega^{(6)}&=&\partial\rho^{(5)}+\frac12 d^{(5)}
\partial (i_k\Lambda) \, .
\end{array}
\end{equation}

Finally, the new gauge parameters in (C.1) and (C.2) are related 
to the eleven-dimensional ones as follows: 

\begin{equation}
{\hat \rho}^{(5)}_{i_1\dots i_4 6}=\rho^{(4)}_{i_1\dots i_4}\, ,
\,\,\,\,\,\,
{\hat \rho}^{(6)}_{i_1\dots i_5 6}=\rho^{(5)}_{i_1\dots i_5}\, .
\end{equation}



\begin{thebibliography}{99}


\bibitem{Romans} L.J.~Romans, 
                   {\sl Massive N=2a Supergravity in Ten Dimensions},
                    {\it Phys.~Lett.~{\bf B169} (1986) 374}.

\bibitem{BDHS} K. Bautier, S. Deser, M. Henneaux and D. Seminara,
               {\sl No cosmological D=11 supergravity},
               {\it Phys. Lett. {\bf B406} (1997) 49},
               {\tt hep-th/9704131}. 

\bibitem{HLW}
        P.S. Howe, N.D. Lambert and P.C. West,
        {\sl A new massive Type IIA supergravity from compactification},
        {\it Phys. Lett. {\bf B416} (1998) 303},
        {\tt hep-th/9707139};

        I.V.~Lavrinenko, H.~L\"u and C.N.~Pope,
        {\sl Fiber bundles and generalized dimensional reductions},
        {\tt hep-th/9710243}. 

\bibitem{BLO} E.~Bergshoeff, Y.~Lozano and T.~Ort\'\i n,
                {\sl Massive Branes},
                {\it Nucl.~Phys.~{\bf B518} (1998) 363},
                 {\tt hep-th/9712115}.

\bibitem{Hull3} C.M. Hull,
                {\sl Massive string theories from M-theory
                 and F-theory},
                {\tt hep-th/9811021}. 
\bibitem{LO}
        Y.~Lozano,
        {\sl Eleven dimensions from the massive D-2-brane},
        {\it Phys. Lett. {\bf B414} (1997) 52},
        {\tt hep-th/9707011};

        T.~Ort\'{\i}n,
        {\sl A note on the D-2-brane of the massive type IIA theory
        and gauged sigma models},
        {\it Phys.~Lett.~{\bf B415} (1997) 39},
        {\tt hep-th/9707113}. 

\bibitem{BEL} E.~Bergshoeff, E.~Eyras and Y.~Lozano,
                {\sl The Massive Kaluza-Klein Monopole},
                {\it Phys.~Lett.~{\bf B430} (1998) 77},
                {\tt hep-th/9802199}.

\bibitem{U1}
        C.M.~Hull,
        {\sl U-duality and BPS spectrum of super Yang-Mills
        theory and M-theory},
        {\it JHEP {\bf 07} (1998) 018},
        {\tt hep-th/9712075}.

\bibitem{U2}
        S.~Elitzur, A.~Giveon, D.~Kutasov and E.~Rabinovici,
        {\sl Algebraic aspects of matrix theory on $T^d$},
        {\it Nucl.~Phys.~{\bf B509} (1998) 122},
        {\tt hep-th/9707217};

        M.~Blau and M.~O'Loughlin,
        {\sl Aspects of U-duality in matrix theory},
        {\it Nucl.~Phys.~{\bf B525} (1998) 182},
        {\tt hep-th/9712047};

        N.A.~Obers, B.~Pioline and E.~Rabinovici,
        {\sl M-theory and U-duality on $T^d$ with gauge
        backgrounds},
        {\it Nucl.~Phys.~{\bf B525} (1998) 163},
        {\tt hep-th/9712084}.
        
\bibitem{U3} N.A. Obers and B. Pioline,
             {\sl U-duality and M-theory},
             {\tt hep-th/9809039}.  

\bibitem{MO}
        P. Meessen and T. Ort\'{\i}n,
        {\sl An SL(2,Z) multiplet of nine-dimensional
        type II supergravity theories},
        {\tt hep-th/9806120}.

\bibitem{Hull}
        C.M.~Hull,
        {\sl Gravitational Duality, Branes and Charges},
        {\it Nucl.Phys. {\bf B509} (1998) 216},
        {\tt hep-th/9705162}.


\bibitem{EJL} E.~Eyras, B.~Janssen and Y.~Lozano,
                {\sl 5-branes, KK-monopoles and T-duality},
                {\it Nucl. Phys. {\bf B531} (1998) 275},
                {\tt hep-th/9806169}. 

\bibitem{BJO} E.~Bergshoeff, B.~Janssen and T.~Ort\'\i n,
                {\sl Kaluza-Klein Monopoles and Gauged Sigma-Models},
                {\it Phys.~Lett.~{\bf B410} (1997) 132},
                {\tt hep-th/9706117}.


\bibitem{varios} Y. Imamura, {\sl Born-Infeld action and 
                 Chern-Simons term from Kaluza-Klein monopole
                 in M-theory},
                 {\it Phys. Lett. {\bf B414} (1997) 242},
                 {\tt hep-th/9706144};

                 A. Sen, {\sl Dynamics of multiple KK-monopoles in
                 M- and string theory}, 
                 {\it Adv. Theor. Math. Phys. {\bf 1} (1998) 115},
                 {\tt hep-th/9707042};

                 A. Hanany and G. Lifschytz,
                 {\sl M(atrix) theory on $T^6$ and a m(atrix) theory
                 description of KK-monopoles},
                 {\tt hep-th/9708037};

                 I. Brunner and A. Karch,
                 {\sl Matrix description of M-theory on $T^6$},
                 {\it Phys. Lett. {\bf B416} (1998) 67},
                 {\tt hep-th/9707259};

                 R. Gregory, J.A. Harvey and G. Moore,
                 {\sl Unwinding strings and T-duality of KK and
                 H monopoles},
                 {\tt hep-th/9708086}.     

\bibitem{HS}
            C.M.~Hull and B.S.~Spence,
            {\sl The geometry of the gauged sigma model with 
            Wess-Zumino term},
            {\it Nucl.~Phys.~{\bf B353} (1991) 379}.
      
        
\bibitem{BRGPT} E.~Bergshoeff, M.~de Roo, M.~B.~Green, 
                  G.~Papadopoulos and P.~K.~Townsend,
                  {\sl Duality of type II 7-Branes and 8-Branes},
                  {\it Nucl.~Phys.~{\bf B470} (1996) 113},
                  {\tt hep-th/9601150}.
 
\bibitem{BREJS}
        E. Bergshoeff, M. de Roo, E. Eyras, B. Janssen and
        J.P. van der Schaar,
        {\sl Intersections involving waves and monopoles in
        eleven dimensions},
        {\it Class. Quant. Grav. {\bf 14} (1997) 2757},
        {\tt hep-th/9704120}. 


\bibitem{BGT}
        E.~Bergshoeff, J.~Gomis and P.K.~Townsend,
        {\sl M-brane intersections from worldvolume superalgebras},
        {\it Phys.~Lett.~{\bf B421} (1998) 109}, 
        {\tt hep-th/9711043}.

\bibitem{deRoo}
              M. de Roo,
              {\sl Intersecting branes and supersymmetry},
              {\tt hep-th/9703124}.

\bibitem{BvdS}
           E.~Bergshoeff and J.P.~van der Schaar, 
           {\sl On M-9-branes},
           {\tt hep-th/9806069}.

\bibitem{BEHHLS}
        E. Bergshoeff, E. Eyras, R. Halbersma, C.M. Hull, Y. Lozano
        and J.P. van der Schaar,
        {\sl Spacetime-filling branes and strings with sixteen
        supercharges}, in preparation.
        

\bibitem{proci}
        E.~Eyras and Y.~Lozano,
        {\sl Brane actions and string dualities},
        Contribution to the proceedings of the TMR-meeting
        {\it Quantum Aspects of Gauge Theories, Supersymmetry and
        Unification}, Corfu, September 1998. 
       

\bibitem{Tsey}
        A.A.~Tseytlin,
        {\sl Harmonic superpositions of M-branes},
        {\it Nucl.~Phys.~{\bf B475} (1996) 149},
        {\tt hep-th/9604035}.

\bibitem{Papa}
        G. Papadopoulos,
        {\sl T-duality and the worldvolume solitons of
        five-branes and KK-monopoles},
        {\it Phys. Lett. {\bf B434} (1998) 277},
        {\tt hep-th/9712162}.

\bibitem{HW}
        A.~Hanany and E.~Witten,
        {\sl Type IIB superstrings, BPS monopoles and three-dimensional
          gauge dynamics},
        {\it Nucl.~Phys.~{\bf B492} (1997) 152},
        {\tt hep-th/9611230}. 

\bibitem{EGJP}
        E.~Bergshoeff, G.~Papadopoulos and J.P.~van der Schaar,
        {\sl Domain-walls on the brane},
        {\it Phys.~Lett.~{\bf B430} (1998) 63},
        {\tt hep-th/9801158}.

\bibitem{sixdi}
        N. Seiberg,
        {\sl A matrix description of M-theory on $T^5$ and
        $T^5/Z_2$},
        {\it Phys. Lett. {\bf B408} (1997) 98},
        {\tt hep-th/9705221};

        E. Witten,
        {\sl New ``gauge'' theories in six dimensions},
        {\it JHEP {\bf 01} (1998) 001},
        {\tt hep-th/9710065}.  

\bibitem{Eric-Mees} E.~Bergshoeff and M.~de Roo,
                {\sl D-branes and T-duality},
                {\it Phys.~Lett. {\bf B380} (1996) 265},
                {\tt hep-th/9603123}.

\bibitem{Green-Hull-Townsend}
                M.B.~Green, C.M.~Hull and P.K.~Townsend,
                {\sl D-p-brane Wess-Zumino actions, T-duality and
                the cosmological constant},
                {\it Phys. Lett. {\bf B382} (1996) 65},
                {\tt hep-th/9604119}.

\bibitem{BHO}
             E.~Bergshoeff, C.M.~Hull and T.~Ort\'{\i}n,
             {\sl Duality in the type II superstring effective
             action},
             {\it Nucl.~Phys.~{\bf B451} (1995) 547},
             {\tt hep-th/9504081}.

\end{thebibliography}
\end{document}